\documentstyle[aps,psfig,multicol]{revtex}
\input{epsf}
\newcommand{\bcols}{\ifpreprintsty\else\begin{multicols}{2}\fi}
\newcommand{\ecols}{\ifpreprintsty\else\end{multicols}\fi}
\begin{document}
\draft
\title{Anomalous power law of quantum reversibility for
classically regular dynamics}
\author{Ph. Jacquod, I. Adagideli, and C.W.J. Beenakker}
\address{Instituut-Lorentz, Universiteit Leiden, P.O. Box 9506, 2300 RA
Leiden, The Netherlands}
\date{\today}
\maketitle
\begin{abstract}
The Loschmidt Echo $M(t)$ (defined as the squared overlap of wave packets
evolving with two slightly different Hamiltonians) is
a measure of quantum reversibility.
We investigate its behavior for classically quasi-integrable systems.
A dominant regime emerges where $M(t) \propto t^{-\alpha}$ 
with $\alpha=3d/2$ depending solely on the dimension $d$ of the system.
This power law decay is faster than the result $\propto t^{-d}$
for the decay of classical phase space densities. 
\end{abstract}
%\narrowtext
\bcols
The search for quantum signatures of chaos has provided much insight
into how classical dynamics manifests itself
in quantum mechanics \cite{Haa00,Gutzwiller}. The basic question is how
to determine from a system's quantum properties whether the classical limit of
its dynamics is chaotic or regular. One very successful approach has been
to look at the spectral statistics, in particular the
distribution of level spacings \cite{Bohigas}. An altogether different
approach, advocated by Schack and Caves \cite{Schack}, has been to 
investigate the sensitivity of the quantum dynamics to perturbations
of the Hamiltonian. This approach goes back to the early work of
Peres \cite{Per84} and has attracted new interest recently in connection
with the study of decoherence and quantum 
reversibility \cite{Jal01,Jac01,echorefs,giulio,Wis01,Prosen02,Seligman02}.

The basic quantity in this approach is the so-called 
Loschmidt Echo, i.e. the fidelity
\begin{equation}\label{fidelity}
M(t)=|\langle\psi_0|\exp(iHt)\exp(-iH_{0}t)|\psi_0\rangle|^{2}
\end{equation}
with which a narrow wavepacket $\psi_0$ can be reconstructed by inverting the
dynamics after a time $t$ with a perturbed Hamiltonian 
$H=H_{0}+V$ \cite{Per84,Jal01}.
(We set $\hbar=1$.)
The fidelity quantifies the sensitivity of the time-reversal 
operation to the uncertainty in the Hamiltonian, and thus provides 
for a measure of quantum reversibility. 

To date, most investigations of $M(t)$ focused on classically chaotic
Hamiltonians $H$ and $H_0$ \cite{Jal01,Jac01,echorefs,giulio,Wis01}. One
notable exception it the original paper by Peres \cite{Per84}, who noted
that the decay of $M(t)$ is slower in a regular system ---
but did not quantify it further. We will show in this article that in a 
regular system a dominant regime
emerges where $M(t)$ has a power law decay $\propto t^{-3d/2}$,
with an exponent depending solely on the dimension $d$ of the system.
This power law decay establishes the higher degree of quantum reversibility 
of regular systems
compared to chaotic ones, where $M(t)$ decays exponentially.
This trend is as expected from classical reversibility (defined in terms
of the decay of the  overlap of classical 
phase space distributions \cite{eckhardt02}). However, we find that 
quantum mechanics plays a crucial role in regular systems by inducing a
parametrically faster power law decay $\propto t^{-3d/2}$ than the 
classical one $\propto t^{-d}$.

We consider the generic situation of a regular or quasi-integrable $H_0$
and a perturbation potential $V$ that has no common integral of motion
with $H_0$. (By regular or quasi-integrable 
we mean systems with a phase space dominated by invariant tori.) 
This condition ensures that, classically, the perturbation has
a component transverse to the invariant tori almost everywhere 
in phase space. Our investigation will 
moreover focus on a regime of sufficiently
strong perturbation (defined below), 
where one expects a fast decay of the 
perturbation correlator. This regime is to be contrasted
with the linear response regime considered in Ref. \cite{Prosen02}. 

We follow the semiclassical approach
of Jalabert and Pastawski \cite{Jal01}. We start from
a Gaussian wavepacket $\psi_0({\bf r}_0') = 
(\pi \sigma^2)^{-d/4} \exp[i {\bf p}_0 \cdot ({\bf r}_0'-{\bf r}_0)-
|{\bf r}_0'-{\bf r}_0|^2/2 \sigma^2]$
and approximate its time evolution by
\begin{eqnarray}\label{propwp}
\exp(-i H t) \psi_0({\bf r}) & = & \int d{\bf r}_0'
\sum_s K_s^H({\bf r},{\bf r}_0';t) \psi_0({\bf r}_0'), \\
K_s^H({\bf r},{\bf r}_0';t) & = & C_s^{1/2} 
\exp[i S_s^H({\bf r},{\bf r}_0';t)-i \pi \mu_s/2].
\end{eqnarray}
\noindent The semiclassical propagator is expressed
as a sum over classical trajectories (labelled $s$)
connecting ${\bf r}$ and ${\bf r}_0'$ in the time $t$.
For each $s$, the partial propagator contains
the action integral $S_s^H({\bf r},{\bf r}_0';t)$ along $s$,
a Maslov index $\mu_s$ (which will drop out), and
the determinant $C_s$ of the monodromy matrix.
Since we consider a narrow initial wavepacket, we linearize
the action in ${\bf r}_0'-{\bf r}_0$
and perform the integration over ${\bf r}_0'$. 
After a 
stationary phase approximation, the semiclassical fidelity reads
\ecols
\begin{eqnarray}\label{semicl}
M(t) & = & (4 \pi \sigma^2)^d \left|\int d{\bf r} 
\sum_s K_s^H({\bf r},{\bf r}_0;t)^* K_s^{H_0}({\bf r},{\bf r}_0;t)  
\exp(-\sigma^2 |{\bf p}_s-{\bf p}_0|^2) \right|^2,
\end{eqnarray}
%\bcols
\noindent with initial momentum
${\bf p}_s=-\partial S_s/\partial {\bf r}_0$.

Eqs. (\ref{propwp}--\ref{semicl}) are equally valid
for regular and chaotic Hamiltonians, as long as semiclassics
applies. Squaring the amplitude in 
Eq.\ (\ref{semicl}) leads to a double sum over classical paths $s$ and $s'$ 
and a double integration over coordinates ${\bf r}$ and
${\bf r}'$. Accordingly, $M(t)=M^{\rm (d)}(t)+M^{\rm (nd)}(t)$
 splits into diagonal ($s=s'$) and nondiagonal ($s \ne s'$) contributions. 
The diagonal contribution sensitively depends
on whether $H_0$ is regular or chaotic. Ref. \cite{Jal01}
found that $M^{\rm (d)}(t) \propto \exp(-\lambda t)$ for chaotic dynamics,
with $\lambda$ the Lyapunov exponent. We will show that
the decay turns into a power law
$M^{\rm (d)}(t) \propto t^{-3d/2}$ for regular dynamics. 
The nondiagonal contribution, on the contrary, is insensitive to the nature
of the classical dynamics (set by $H_0$), provided
the perturbation Hamiltonian $V$ has no common integral of motion
with $H_0$.
Ref. \cite{Jac01} found that $M^{\rm (nd)}(t) \propto \exp(-\Gamma t)$
for chaotic dynamics, with $\Gamma$ 
given by the golden rule spreading width of an eigenstate
of $H_0$ over the eigenbasis of $H$. (This golden rule decay requires
that $\Gamma $ is larger than the level spacing $\Delta$, but smaller
than the bandwidth.)
We will see that the same exponential decay of $M^{\rm(nd)}(t)$
holds when $H_0$ is regular, so that $M^{\rm (d)}(t)$ always dominates
in the long time limit.
Consequently, the fidelity decays exponentially,
$\propto \exp[-{\rm min}(\Gamma,\lambda) t]$ 
for chaotic systems, while for regular systems the decay is 
algebraic, $\propto t^{-3d/2}$, 
as it is then set by the diagonal contribution. 
The golden rule width $\Gamma$ still determines the regime of validity of 
the power law decay via the condition $\Gamma > \Delta$.

Continuing from Eq. (\ref{semicl}), and still following Ref. \cite{Jal01},
we write $M(t)$ as
%\ecols
\begin{eqnarray}\label{mtot}
M(t) & = & (4 \pi \sigma^2)^d \int d{\bf r} \int d{\bf r}'
\sum_{s,s'} C_s C_{s'} \exp[i \delta S_s({\bf r},{\bf r}_0;t) - 
i \delta S_{s'}({\bf r}',{\bf r}_0;t) ] 
\mbox{} \exp(-\sigma^2 |{\bf p}_s-{\bf p}_0|^2-\sigma^2 
|{\bf p}_{s'}-{\bf p}_0|^2),
\end{eqnarray}
\noindent with $ \delta S_s({\bf r},{\bf r}_0;t) =
S_s^H({\bf r},{\bf r}_0;t)-S_s^{H_0}({\bf r},{\bf r}_0;t)$.
Considering first the diagonal contribution $M^{\rm (d)}(t)$,
we set $s = s'$ and expand the phase difference as
\begin{equation}
\delta S_s({\bf r},{\bf r}_0;t) - 
\delta S_s({\bf r}',{\bf r}_0;t) = \int_0^t 
d{\tilde t} \ \nabla V[{\bf q}({\tilde t})] \cdot
\bigl({\bf q}(\tilde{t}\;) - {\bf q}'(\tilde{t}\;)\bigr).
\end{equation}
The points ${\bf q}$ and ${\bf q}'$ lie on the classical
path with ${\bf q}(t)={\bf r}$, ${\bf q}'(t)={\bf r}'$, and
${\bf q}(0)={\bf q}'(0)={\bf r}_0$. 
In a regular system, the distance between two initially close
points increases linearly with time, 
$|{\bf q}(\tilde{t}) - {\bf q}'(\tilde{t})| \simeq (\tilde{t}/t)
|{\bf r}-{\bf r}'|$. Here we depart from the exponential divergence
$\propto \exp[\lambda(\tilde{t}-t)]$ assumed in Ref. \cite{Jal01}
for chaotic dynamics.

The spatial integrations and the sums over classical paths
in Eq. (\ref{mtot}) lead to the phase averaging
\begin{equation}\label{clt}
\exp(i \delta S_s - i \delta S_{s}') \rightarrow 
\langle \exp(i \delta S_s - i\delta S_{s}') \rangle \simeq \exp[-\case{1}{2}
\langle (\delta S_s-\delta S_s')^2 \rangle ].
\end{equation}
\noindent Since $V$ and $H_0$ have no common integral
of motion, we may expect a fast decay of the correlations,

\begin{equation}\label{correl}
\langle \partial_i V[{\bf q}(\tilde{t})] \partial_j 
V[{\bf q}(\tilde{t}')] \rangle = U \delta_{ij} 
\delta({\tilde t}-{\tilde t}'). 
\end{equation}
One then gets
\begin{eqnarray}
M^{\rm (d)}(t) & = & (4 \pi \sigma^2)^d \int d{\bf r} \int d{\bf r}' \sum_s 
C_s^2 \exp\bigl(-\case{1}{2} U\int_0^t d {\tilde t} \ (\tilde{t}/t)^2
|{\bf r}-{\bf r}'|^2 \bigr) 
\exp(-2 \sigma^2 |{\bf p}_s-{\bf p}_0|^2 )\nonumber \\
& = & (4 \pi \sigma^2)^d \int d{\bf r}_+ \int d{\bf r}_- \sum_s C_s^2
\exp\bigl(-\case{1}{6} U t \; {\bf r}_-^2\bigr) 
\exp(-2 \sigma^2 |{\bf p}_s-{\bf p}_0|^2).
\end{eqnarray}
\noindent The Gaussian integration over ${\bf r}_- \equiv 
{\bf r}-{\bf r}'$ ensures that ${\bf r} \approx {\bf r}'$, 
and hence ${\bf r}_+ \equiv ({\bf r}+{\bf r}')/2 \approx {\bf r}$.
One $C_s$ is then absorbed by a change of variable from ${\bf r}_+$
to ${\bf p}_s$, and the Gaussian integral over ${\bf r}_-$ gives a 
factor $\propto t^{-d/2}$.
Finally, setting $C_s \approx t^{-d}$ as is the case in a regular
system, we arrive at 
\begin{equation}\label{diagdecay}
M^{\rm (d)}(t) \propto t^{-3d/2},
\end{equation} 
\noindent which is the central result of this paper. 
The power law (\ref{diagdecay}) holds
once the perturbation is strong enough
to induce a golden rule spreading of the eigenstates of $H_0$
over the eigenbasis of $H$ (which is the range of validity \cite{Jal01,Jac01} 
of the above semiclassical approach), and under the 
assumption that the perturbation potential varies rapidly
along a classical trajectory of $H_0$. [We used this assumption
to average the complex exponential in Eq.\ (\ref{clt}).] 
The decay exponent ${3d/2}$ is insensitive to the
choice (\ref{correl}) 
of a $\delta$-function force correlator. Even a power-law decaying correlator 
$\propto |{\tilde t}-{\tilde t}'|^{-\alpha}$ (with $\alpha \ge 1$)
results in the same exponent as in Eq.\ (\ref{diagdecay}).

The nondiagonal contribution ($s \ne s'$) to Eq.\ (\ref{mtot}) is the same
as in Refs.\ \cite{Jal01,Jac01}. The phase averaging 
can be performed separately for $s$ and $s'$ and one gets
\begin{equation}\label{phase}
\langle \exp[i \delta S_s] \rangle = 
\exp(-\case{1}{2} \langle \delta S_s^2 \rangle ) = \mbox{}
\exp\left(-\case{1}{2} \int_0^t d{\tilde t} \int_0^t d{\tilde t}' 
\langle V[{\bf q}(\tilde{t})]
V[{\bf q}(\tilde{t}')] \rangle \right).
\end{equation}
\bcols \noindent The point ${\bf q}(\tilde{t})$ lies on path $s$ with 
${\bf q}(0)={\bf r}_0$ and ${\bf q}(t)={\bf r}$.
If $V$ and $H_0$ have no common integral of motion, 
the correlator of $V$ gives the golden rule decay $\propto \exp(-\Gamma t)$
regardless of whether $H_0$ is chaotic or regular \cite{caveat2}. 
We conclude that for regular systems, the fidelity is
dominated by the algebraically decaying diagonal contribution.

In order to check numerically the analytical result 
(\ref{diagdecay}), we have studied the kicked top Hamiltonian \cite{Haa00}
\begin{eqnarray}
H_{0}=(\pi/2\tau)S_{y}+(K/2S)S_{z}^{2}\sum_{n}\delta(t-n\tau), 
\label{H0def}
\end{eqnarray}
\noindent which describes a vector spin of conserved 
magnitude $S$, undergoing a free precession
around the $y$-axis, which is periodically
perturbed (period $\tau$) by sudden rotations
around the $z$-axis over an angle proportional to $S_{z}$.
Because $S$ is conserved,
$H_0$ is a one-dimensional Hamiltonian ($d=1$), with a two-dimensional
classical phase space consisting of 
the sphere of radius $S=1$. The canonically conjugated variables 
are $(\varphi,\cos \theta)$, where
$\theta$ and $\varphi$ are spherical coordinates.

The classical limit of the kicked top is given by the map \cite{Haa00}
\begin{eqnarray}\label{clmap}
\cases{
x_{n+1} = z_n \cos(K x_n) + y_n \sin(K x_n) \cr
y_{n+1} = -z_n \sin(K x_n) + y_n \cos(K x_n) \cr 
z_{n+1} = -x_n, \cr}
\end{eqnarray}
in the cartesian coordinates $x=\sin \theta \cos \varphi$, 
$y= \sin \theta \sin \varphi$, and $z=\cos \theta$.
Depending on the kicking strength $K$, the classical dynamics is regular,
partially chaotic, or fully chaotic. 
We consider a kicking strength $K=1.1$ for which 
the dynamics is regular for most of phase space.
We checked that our results are not sensitive to the value 
of $K$, as long as the dynamics remains regular.

The quantum mechanical time evolution after $n$ periods is given 
by the $n$-th power of the Floquet operator
\begin{equation}
F_{0}=\exp[-i(K/2S)S_{z}^{2}]\exp[-i(\pi/2)S_{y}]. \label{Fdef}
\end{equation}
We perturb the reversed time evolution by a
periodic rotation of constant angle around the $x$-axis, slightly
delayed with respect to the kicks in $H_0$,
\begin{equation}
V=\phi S_{x}\sum_{n}\delta(t-n\tau-\epsilon). \label{H1def}
\end{equation}
The corresponding Floquet operator is $F=\exp(-i\phi S_{x})F_{0}$. 
We set $\tau=1$ for ease of notation, and varied $S$
between $250$ and $1000$
(both $H$ and $H_{0}$ conserve the spin magnitude). We calculated the
average decay $\overline{M}$ of 
$M(t=n)=|\langle\psi_0|(F^{\dagger})^{n}F_{0}^{n}|\psi_0\rangle|^{2}$ 
taken over 50 to 200 initial Gaussian wavepackets $\psi_0$ of minimal
spreading (coherent states).

In Fig.\ 1 we show the decay of $\overline{M}$ 
for $S=1000$ and different perturbation strengths $\phi$. For
weak perturbations, the decay of  $\overline{M}$ is exponential, and not 
Gaussian as one would expect from first order perturbation 
theory \cite{Per84}. The reason why the perturbation operator $S_x$ gives 
no first order correction is that for $K=1.1$,
eigenfunctions of $F_0$ are almost identical to eigenfunctions of $S_y$,
so that diagonal matrix elements of $V$ vanish in this basis. 
For weak $\phi$, the local spectral density of states $\rho(\epsilon)$ 
consists then of a delta function at zero energy plus an algebraically
decaying tail \cite{Cohen00}. Because of the absence of
a first-order correction, the decay of the fidelity
is given by the Fourier transform of this tail \cite{Wis01}. 
We numerically obtained a decay 
$\rho(\epsilon) \propto (\epsilon^2+\gamma^2/4)^{-1}$ 
with $\gamma \propto \phi^{1.5}$. The resulting 
exponential decay $\propto \exp(-\gamma t)$ of the fidelity 
differs from the golden
rule decay $\propto \exp(-\Gamma t)$ with $\Gamma \propto \phi^2$.
As $\phi$ increases, the decay of $\overline{M}$ turns into the predicted  
power law $\propto t^{-3/2}$, which prevails as soon as one
enters the golden rule regime, i.e. for 
$\Gamma/\Delta \approx \phi^2 S^3 \gtrsim 1$ \cite{Jac01}. 
One therefore expects the power law
decay to appear as $S$ is increased at fixed $\phi$,
which is indeed observed in the inset to Fig.\ 1.

We checked that these results are not sensitive to our choice of
Hamiltonian, by replacing $S_x$ in Eq.\ (\ref{H1def}) with 
$S_z^2$ (this is the model used in Ref.\ \cite{Prosen02}) and also by
studying a kicked rotator as an alternative model to the kicked top.
These numerical results all give
clear confirmation of the power law decay (\ref{diagdecay}).

It is instructive to contrast these results for the decay of the overlap
of quantum wavefunctions with the decay of the overlap of classical 
phase space distributions, a ``classical fidelity'' problem that has
recently been investigated \cite{giulio,Prosen02,eckhardt02}. We assume that
the two phase space distributions $\rho_0$ and $\rho$ are initially
identical and evolve according to the Liouville equation of
motion corresponding to the classical map (\ref{clmap})
for two different Hamiltonians $H_0$ and $H$. We consider regular
dynamics and ask for the decay of the normalized phase space 
overlap 
\begin{equation}\label{classic}
M_c(t)=\int d{\bf x}\int d{\bf p} \; \rho_0({\bf x},{\bf p};t) 
\; \rho({\bf x},{\bf p};t)/N_{\rho},
\end{equation}
where $N_{\rho}=(\int d{\bf x}\int d{\bf p} \; \rho_0)^{1/2}
(\int d{\bf x}\int d{\bf p} \; \rho)^{1/2}$.

We have found above that a factor $\propto t^{-d/2}$ in the decay
of the quantum fidelity $M(t) \propto t^{-3d/2}$ originates from the action 
phase difference and is thus of purely quantum origin.
One therefore expects a slower 
classical decay $M_c(t) \propto C_s \propto t^{-d}$.
In Fig.\ 2 we show the decay of the averaged $\overline{M}_c$ taken over 
$10^4$ initial points within a narrow volume of phase space
$\sigma \equiv \sin \theta \delta \theta \; \delta \varphi$, for $K=1.1$ and
$\phi=1.7 \cdot 10^{-4}$. The decay is
$\overline{M}_c \propto t^{-1}$, and clearly differs 
from the quantum decay $\propto t^{-3/2}$. 

The power law decay prevails
for classically weak perturbations, for which the center of mass 
of $\rho$ and $\rho_0$ stay close together. [This is required by
the stationary phase condition leading to Eq. (4).]
Keeping $\sigma$
fixed, and increasing the perturbation strength $\phi$, the invariant
tori of $H_0$ start to differ significantly from those of $H$ 
on the resolution scale $\sigma$, giving a threshold $\phi_c \approx \sigma$.
Above $\phi_c$, the distance 
between the center of mass of $\rho_0$ and $\rho$ increases 
with time $\propto t$ and one
expects a much faster decay $M_c(t) \propto \exp[-{\rm const}\times t^2]$ for 
classical Gaussian phase space distributions \cite{eckhardt02}. 
Quantum mechanically, $\sigma=1/S$ (the effective Planck
constant) and the threshold translates into
$\phi_c \sim 1/S$, coinciding with the upper boundary of the 
golden rule regime. As long as one stays in that regime, the perturbation
will affect the phase in Eq. (7), and result in the anomalous power law
decay $\propto t^{-3d/2}$. 

In conclusion, our investigations of the Loschmidt Echo
(\ref{fidelity}) in the generic regime of classically quasi-integrable 
dynamics show that its decay is dominated by the power law 
$M(t) \propto t^{-\alpha}$. While from purely classical considerations
one expects an exponent $\alpha_c=d$, we semiclassically obtain
an anomalous exponent $\alpha=3d/2$. This is
corroborated by numerical simulations. The power law
decay is to be contrasted with the exponential decay found
for chaotic systems, thereby providing for a novel ``quantum signature
of chaos''.

This work was supported by the Dutch Science Foundation NWO/FOM and 
the U.S. Army Research Office (grant DAAD 19-02-0086). 
We thank B. Eckhardt, T. Prosen, and T. Seligman
for useful remarks.

\begin{figure}
\begin{center}
\epsfxsize=3.3in
\epsffile{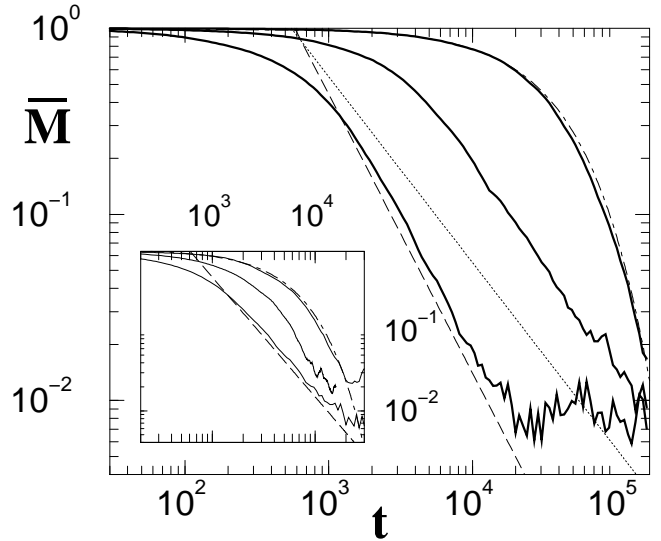}
\end{center}
\vspace{-0.2cm}
\caption{
Decay of $\overline{M}$ for $S=1000$, 
$K=1.1$, and $10^{5} \,\phi=1.5$, 4.5, and 10 (thick
solid lines from right to left). The crossover from exponential
to power-law decay is illustrated by the dotted-dashed line 
$\propto \exp[-2.56 \cdot 10^{-5} \,t]$ and the dashed 
line $ \propto t^{-3/2}$. The dotted line gives the classical
decay $\propto t^{-1}$.
Inset: Decay of $\overline{M}$ for $K=1.1$,
$\phi=10^{-4}$, and $S=250$, $500$, and $1000$ (solid lines from right 
to left). The dashed and dotted-dashed lines indicate
the power law $\propto t^{-3/2}$ and exponential 
$\propto \exp[-2 \cdot 10^{-4} \,t]$ decay, respectively.
These plots show that the $t^{-3/2}$ decay is reached either by
increasing the perturbation strength $\phi$ at fixed spin
magnitude $S$, or by increasing $S$ at fixed $\phi$.}
\label{fig:fig1}
\end{figure}

\begin{figure}
\epsfxsize=3.3in
\begin{center}
\vspace{-1cm}
\epsffile{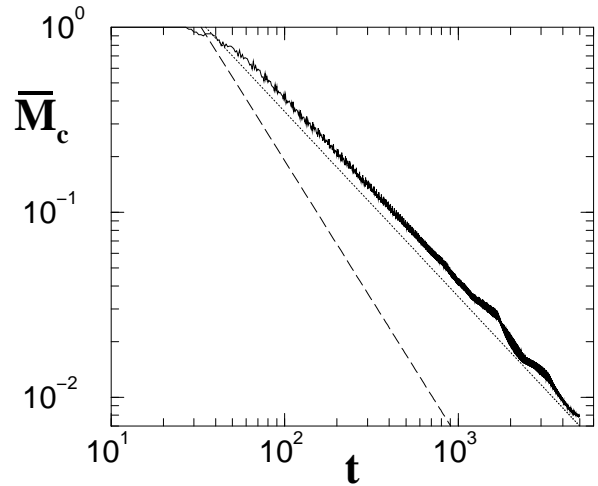}
\end{center}
\vspace{-0.5cm}
\caption
{Decay of the average overlap (\ref{classic})
of classical phase space distributions, for the kicked top with $K=1.1$
and $\phi=1.7 \cdot 10^{-4}$ (solid line). The dotted and dashed lines
give the classical and quantum 
power law decays $\propto t^{-1}$ and $ \propto t^{-3/2}$,
respectively.}
\label{fig:fig2}
\end{figure}

\ecols

\begin{references}
\bibitem{caveat2} This conclusion, that the golden rule decay 
holds whether $H_0$ is regular or chaotic, 
can also be obtained via a fully quantum mechanical 
approach based on random-matrix theory assumptions for $V$.
The invariance under unitary transformations of the distribution
of $V$ is sufficient to obtain the exponential decay 
$M^{\rm (nd)}(t) \propto \exp(-\Gamma t)$, irrespective of the
distribution of $H_0$.
\end{references}
\end{document}